\documentclass[preprint]{aastex}
\usepackage{amsmath}
\usepackage{natbib}
\usepackage{ulem}
\bibpunct{(}{)}{;}{a}{,}{,}

\shorttitle{Spectral Anisotropy of Els\"asser Variables in Two Dimensional Wave-vector Space}
\shortauthors{Yan et al.}

\begin{document}


\title{Spectral Anisotropy of Els\"asser Variables in Two Dimensional Wave-vector Space as Observed in the Fast Solar Wind Turbulence}

\author{Limei Yan\altaffilmark{1}, Jiansen He\altaffilmark{1}, Lei Zhang\altaffilmark{1}, Chuanyi Tu\altaffilmark{1}, Eckart Marsch\altaffilmark{2}, Christopher H.K. Chen\altaffilmark{3}, Xin Wang\altaffilmark{1}, Linghua Wang\altaffilmark{1}, Robert T. Wicks\altaffilmark{4}}

\altaffiltext{1}{School of Earth and Space Sciences, Peking University, 100871 Beijing, China, jshept@gmail.com} \altaffiltext{2}{Institute for Experimental and Applied Physics, Christian Albrechts University at Kiel, D-24118 Kiel, Germany} \altaffiltext{3}{Department of Physics, Imperial College London, London, SW7 2AZ, United Kingdom} \altaffiltext{4}{Department of Space and Climate Physics, University College London, Gower Street, London, UK}

\begin{abstract}
Intensive studies have been conducted to understand the anisotropy of solar wind turbulence. However, the anisotropy of Els\"asser variables ($\textbf{\textbf{\textrm{Z}}}^{\pm}$) in 2D wave-vector space has yet to be investigated. Here we first verify the transformation based on the projection-slice theorem between the power spectral density $\textrm{PSD}_{\textrm{2D}}(k_{\parallel},k_{\perp} )$ and the spatial correlation function $\textrm{CF}_{\textrm{2D}} (r_{\parallel},r_{\perp} )$.
Based on the application of the transformation to the magnetic field and the particle measurements from the WIND spacecraft, we investigate the spectral anisotropy of Els\"asser variables ($\textbf{\textrm{Z}}^{\pm}$), and the distribution of residual energy $\textrm{E}_{\textrm{R}}$, Alfv\'en ratio $\textrm{R}_{\textrm{A}}$ and Els\"asser ratio $\textrm{R}_{\textrm{E}}$ in the $(k_{\parallel},k_{\perp} )$ space. The spectra $\textrm{PSD}_{\textrm{2D}}(k_{\parallel},k_{\perp} )$ of \textbf{B}, \textbf{V}, and $\textbf{\textrm{Z}}_{\textrm{major}}$ (the larger of $\textbf{\textrm{Z}}^{\pm}$) show a similar pattern that $\textrm{PSD}_{\textrm{2D}}(k_{\parallel},k_{\perp} )$ is mainly distributed along a ridge inclined toward the $k_{\perp}$ axis. This is probably the signature of the oblique Alfv\'enic fluctuations propagating outwardly. Unlike those of \textbf{B}, \textbf{V}, and $\textbf{\textrm{Z}}_{\textrm{major}}$, the spectrum $\textrm{PSD}_{\textrm{2D}}(k_{\parallel},k_{\perp} )$ of $\textbf{\textrm{Z}}_{\textrm{minor}}$ is distributed mainly along the $k_{\perp}$ axis. Close to the $k_{\perp}$ axis, $\left |\textrm{E}_{\textrm{R}}\right|$ becomes larger while $\textrm{R}_{\textrm{A}}$ becomes smaller, suggesting that the dominance of magnetic energy over kinetic energy becomes more significant at small $k_{\parallel}$. $\textrm{R}_{\textrm{E}}$ is larger at small $k_{\parallel}$, implying that $\textrm{PSD}_{\textrm{2D}}(k_{\parallel},k_{\perp} )$ of $\textbf{\textrm{Z}}_{\textrm{minor}}$ is more concentrated along the $k_{\perp}$ direction as compared to that of $\textbf{\textrm{Z}}_{\textrm{major}}$. The residual energy condensate at small $k_{\parallel}$ is consistent with simulation results in which $\textrm{E}_{\textrm{R}}$ is spontaneously generated by  Alfv\'en wave interaction.

\end{abstract}

\keywords{solar wind --- turbulence --- waves}

\section{Introduction}

MHD turbulence in the solar wind is considered to be a cascade of energy over different scales caused by the nonlinear interaction between counterpropagating Alfv\'{e}n waves, which has been studied in detail by asymptotic solution \citep{Howes2013PhPl1} and numerical simulation \citep{Nielson2013PhPl2}. The cascade is anisotropic with the cascading direction mainly perpendicular to the local mean magnetic field \citep[e.g.][]{GS1995ApJ}. When the oppositely directed Alfv\'{e}n waves carry unequal energy, the turbulence is imbalanced. Imbalanced weak \citep{Galtier2000JPlPh,Lithwick2003ApJ} and strong \citep{Lithwick2007ApJ} turbulence have been studied intensively. In some theoretical studies, the energy spectrum of the Els\"asser variables ($\textrm{Z}^{\pm}=\textrm{\textbf{V}} \pm \textrm{\textbf{b}}$ , $\textrm{\textbf{b}}=\frac{ \textrm{\textbf{B}}}{\sqrt{\mu_{0}\rho}}$) $\textrm{E}^{\pm}$ have a same scaling with different amplitudes.
The scaling is $\textrm{E}^{+} \propto \textrm{E}^{-} \propto k_{\perp}^{-3/2} $  with the phenomenon of dynamic alignment \citep{Perez2009PhRvL}, while the scaling is $\textrm{E}^{+} \propto \textrm{E}^{-} \propto k_{\perp}^{-5/3} $ without the phenomenon of dynamic alignment \citep{Lithwick2007ApJ}. In the solar wind, especially in fast streams, imbalanced turbulence is usually observed (one of $\textbf{\textrm{Z}}^{\pm}$ is dominating). We define the dominant mode as $\textbf{\textrm{Z}}_{\textrm{major}}$ which is typically the Alfv\'en wave propagating outward from the sun, while the subdominant mode $\textbf{\textrm{Z}}_{\textrm{minor}}$ is weak and complicated. The subdominant mode has been suggested to be inward propagating Alfv\'en wave at high frequency and compressive events at low frequency \citep[e.g.][]{Bruno1996AIPC}, magnetic structures \citep{Tu1992sws,Tu1993JGR}.

Without temperature anisotropies and relative drifts, if the MHD turbulence is only composed of counterpropagating Alfv\'{e}n waves with no nonlinear interaction, the residual energy $\textrm{E}_{\textrm{R}}=\upsilon^{2}-b^{2}$ would be zero. However in the solar wind, outward propagating Alfv\'en waves are often observed, while inward propagating Alfv\'en waves are rarely observed. Besides, there are also many structures like tangential discontinuities in the solar wind which may contribute more to the magnetic disturbances than the velocity fluctuations. These factors would lead to the residual energy being nonzero. Residual energy is high at low frequency / small $k$ from observations \citep[e.g.][]{Roberts1987JGR, Bavassano1998JGR, Wicks2011PhRvL}, however it remains unknown whether the residual energy is mainly distributed along $k_{\parallel}$ or $k_{\perp}$. In the solar wind turbulence, the residual energy at small scales near the dissipation range is usually less than 0 \citep[e.g.][]{Belcher1971JGR,Matthaeus1982JGR,Boldyrev2012AIPC,Chen2013ApJ}. This is also noted in simulations \citep[e.g.][]{Grappin1983A&A,Muller2005PhRvL,Gogoberidze2012PhPl}. As revealed from recent simulations, the residual energy is concentrated at small $k_{\parallel}$ \citep{Boldyrev2009PhRvL,Wang2011ApJ}.
The distribution of the residual energy in the wave-vector space will allow us to compare with these simulation results.

The presence of a mean magnetic field may lead to spectral anisotropy of magnetohydrodynamic (MHD) turbulence \citep{Shebalin1983JPlPh}. \citet{GS1995ApJ} investigated the anisotropy in a balanced strong MHD turbulence with vanishing cross-helicity, revealing a spectrum perpendicular to the magnetic field of $\textrm{E}(k_{\perp}) \sim k_{\perp}^{-5/3}$, a parallel spectrum $\textrm{E}(k_{\parallel}) \sim k_{\parallel}^{-2}$, and a scaling relation $k_{\parallel} \sim k_{\perp}^{2/3}$ based on the critical balance assumption, i.e., the linear wave periods are comparable to the nonlinear turnover timescales. The anisotropic power and scaling of magnetic field fluctuations in the inertial range of high-speed solar wind turbulence is first reported by \citet{Horbury2008PhRvL}, who introduced the method to estimate the scale-dependent local $\vec{B_{0}}$. The reduced spectrum has an index near -2 when $\theta_{BV}\rightarrow 0$ and an index near $-5/3$ when $\theta_{BV}\rightarrow 90$ where $\theta_{BV}$ is the angle between the magnetic field and the flow. \citet{Podesta2009ApJ} gave similar results using magnetic field measurements from the STEREO. \citet{Luo2010ApJ} and \citet{Chen2011MNRAS} also got a similar conclusion for the magnetic structure function. When the second order structure function of the magnetic fluctuations is decomposed into the components perpendicular ($\delta \textrm{B}_{\perp}^{2}$) and parallel ($\delta \textrm{B}_{\parallel}^{2}$) to the mean field, both components show spectral index anisotropy between the ion and electron gyroscales in the fast solar wind \citep{Chen2010PhRvL}. At these small scales the spectral index of  $\delta \textrm{B}_{\perp}^{2}$ is -2.6 at large angles and -3 or steeper at small angles. This kind of  spectral anisotropy of solar wind turbulence in the inertial range is probably related to the intermittency \citep{Wang2014ApJ}. \citet{Wicks2011PhRvL} studied the anisotropy of the Els\"asser variables in fast solar wind based on the reduced spectrum, finding that the dominant Els\"asser  mode is isotropic at low frequencies but becomes increasingly anisotropic at higher frequencies while the subdominant mode is anisotropic throughout. This result suggests that the anisotropy of the subdominant mode may be stronger than the dominant mode.

The spectral anisotropy has been studied extensively based on the reduced spectrum, while the anisotropy in wave-vector space is relatively rarely studied. The K-filtering method has been applied to the Cluster observations to investigate the anisotropy in wave-vector space \citep[e.g.][]{Sahraoui2010PhRvL,Narita2010PhRvL}. However, this method is sensitive only to a limited number of wave modes and the scales comparable to the inter-spacecraft distance \citep{Horbury2012SSRv}. Based on single spacecraft measurements, \citet{He2013ApJ} first constructed the normalized power spectral density (PSD) of magnetic field fluctuations (\textbf{B}) in 2D wave-vector space. They found that the PSD of \textbf{B} shows an anisotropic distribution, which is mainly characterized by a ridge distribution inclined more toward $k_{\perp}$ as compared to $k_{\parallel}$. The spectral anisotropy of velocity and Els\"asser variables in wave-vector space has not been previously investigated. We will study them in this paper using the method contributed by \citet{He2013ApJ}. Moreover, we will investigate the distribution of residual energy $\textrm{E}_{\textrm{R}}=\textrm{E}_{\textrm{v}}-\textrm{E}_{\textrm{b}}$, Alfv\'en ratio $\textrm{R}_{\textrm{A}}=\frac{\textrm{E}_{\textrm{v}}}{\textrm{E}_{\textrm{b}}}$ and Els\"asser ratio $\textrm{R}_{\textrm{E}}=\frac{\textrm{E}_{\textrm{Z}_{\textrm{minor}}}}{\textrm{E}_{\textrm{Z}_{\textrm{major}}}}$ in the wave-vector space.

\section{Benchmark test of the conversion between $\textrm{CF}_{\textrm{2D}}$ and $\textrm{PSD}_{\textrm{2D}}$}

To test the conversion between $\textrm{CF}_{\textrm{2D}}$ and $\textrm{PSD}_{\textrm{2D}}$ based on the projection-slice theorem, we first assume a double Gaussian distribution, a strong parallel component and a weak perpendicular component, for $\textrm{CF}_{\textrm{2D}}$ using the formula given below:
\begin{equation}
\begin{split}
\label{eq:1}
\textrm{CF}_{\textrm{2D}} \left ( r_{\parallel},r_{\perp} \right )=\textrm{exp}\left(-\frac{r_{\parallel}^{2}}{2\sigma_{\parallel1}^{2}}\right) \cdot \textrm{exp}\left(-\frac{r_{\perp}^{2}}{2\sigma_{\perp1}^{2}}\right)+3 \cdot \textrm{exp}\left(-\frac{r_{\parallel}^{2}}{2\sigma_{\parallel2}^{2}}\right) \cdot \textrm{exp}\left(-\frac{r_{\perp}^{2}}{2\sigma_{\perp2}^{2}}\right)
\end{split}
 \end{equation}
Based on this assumption, there are three ways to obtain the $\textrm{PSD}_{\textrm{2D}}$.
The first way is to get the $\textrm{PSD}_{\textrm{2D}}$ directly from the corresponding formula :
\begin{equation}
\begin{split}
\label{eq:2}
\textrm{PSD}_{\textrm{2D}}\left ( k_{\parallel},k_{\perp} \right )=\sigma_{\parallel1}\cdot \textrm{exp}\left(-\frac{k_{\parallel}^{2}\cdot \sigma_{\parallel1}^{2}}{2}\right) \cdot \sigma_{\perp1} \cdot \textrm{exp}\left(-\frac{k_{\perp}^{2}\cdot \sigma_{\perp1}^{2}}{2}\right)\\+3 \cdot \sigma_{\parallel2}\cdot \textrm{exp}\left(-\frac{k_{\parallel}^{2}\cdot \sigma_{\parallel2}^{2}}{2}\right) \cdot \sigma_{\perp2} \cdot \textrm{exp}\left(-\frac{k_{\perp}^{2}\cdot \sigma_{\perp2}^{2}}{2}\right)
\end{split}
 \end{equation}
The second way is to do the transformation based on the projection-slice theorem. Firstly, we make the 1 dimensional integration (1D-INT) of $\textrm{CF}_{\textrm{2D}}$ along the direction ($\textbf{u}'$) normal to $\textbf{k}$ to get 1D-CF at each angle using the following formula:
\begin{equation}
\begin{split}
\label{eq:3}
\textrm{CF}_{\textrm{1D}}\left ( r,\theta_{r}\right)=\int_{-\infty }^{+\infty }\textrm{CF}_{\textrm{2D}}\left ( r\cos\theta_{k}-{u}'\sin\theta_{k}, r\sin\theta_{k}+{u}'\cos\theta_{k}\right )d{u}'
\end{split}
 \end{equation}
Secondly, we calculate the Fourier transformation (FT) of the 1D-CF to get the corresponding slice of 2D-PSD at each angle using the following formula:
\begin{equation}
\begin{split}
\label{eq:4}
\textrm{PSD}_{\textrm{2D}}\left ( k,\theta_{k}\right)=\int_{-\infty }^{+\infty } \textrm{CF}_{\textrm{1D}} \left ( r,\theta_{k} \right )\textrm{exp}\left(-ikr\right) dr
\end{split}
 \end{equation}
Finally, the $\textrm{PSD}_{\textrm{2D}}$ is assembled by putting the slices of 2D-PSD at each angle together.
The third way is to do the two dimensional Fourier transformation (2D-FT) of $\textrm{CF}_{\textrm{2D}}\left ( r_{\parallel},r_{\perp} \right )$:
\begin{equation}
\begin{split}
\label{eq:5}
\textrm{PSD}_{\textrm{2D}}\left (k_{\parallel},k_{\perp}\right)=\int_{-\infty }^{+\infty }\int_{-\infty }^{+\infty } \textrm{CF}_{\textrm{2D}} \left (r_{\parallel},r_{\perp}\right) \textrm{exp}\left(-i\left ( k_{\parallel}r_{\parallel} +k_{\perp}r_{\perp}\right )\right)dr_{\parallel}dr_{\perp}
\end{split}
 \end{equation}

\begin{figure*}
\centering
\includegraphics[width=14cm]{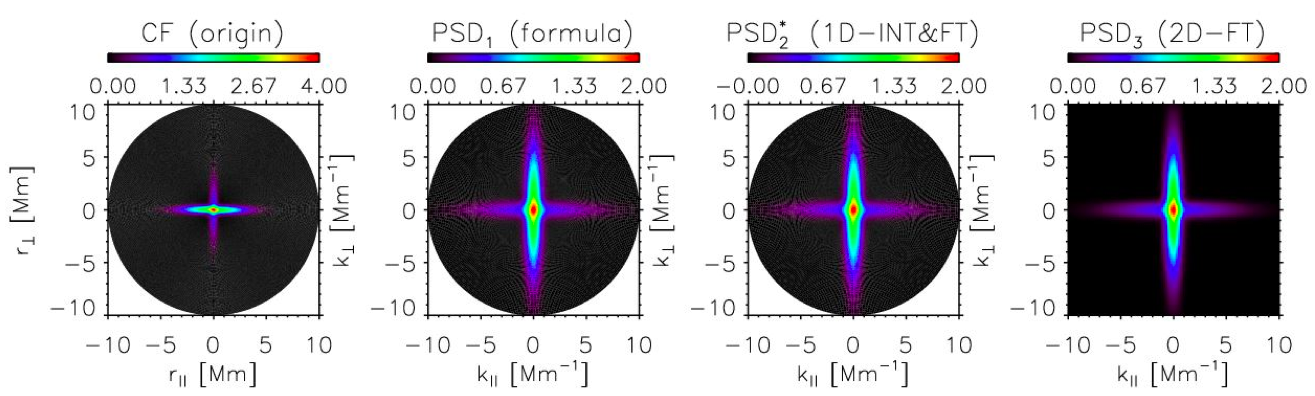}
\caption{Benchmark test of the conversion from $\textrm{CF}_{\textrm{2D}}$ to $\textrm{PSD}_{\textrm{2D}}$. From left to right: the original $\textrm{CF}_{\textrm{2D}}$, $\textrm{PSD}_{\textrm{2D}}$ calculated from Equation \ref{eq:2}, $\textrm{PSD}_{\textrm{2D}}$ obtained from the transformation based on the projection theorem which involves 1D-INT and FT, and $\textrm{PSD}_{\textrm{2D}}$ obtained by 2D-FT.} \label{fig1}
\end{figure*}

 Here, we set $\sigma_{\parallel1}=0.25$, $\sigma_{\perp1}=2.0$, $\sigma_{\parallel2}=2.0$, and $\sigma_{\perp2}=0.25$. The origin $\textrm{CF}_{\textrm{2D}}$ and the transferred $\textrm{PSD}_{\textrm{2D}}$ obtained by the three methods above are given in Figure~\ref{fig1}. From Figure~\ref{fig1}, the $\textrm{PSD}_{\textrm{2D}}$ obtained by the three different ways are in accordance with each other. This confirms that the the conversion between $\textrm{CF}_{\textrm{2D}}$ and $\textrm{PSD}_{\textrm{2D}}$ based on the projection-slice theorem is credible.

\section{Data analysis and results}
Four fast solar wind streams , with their magnetic field measured by the Magnetic Field Investigation (MFI; \citet{Lepping1995SSRv}) and particle distribution measured by the Three-Dimensional Plasma Analyser (3DP; \citet{Lin1995SSRv}), are investigated at the time cadence of 3\,s. The time intervals for the four fast solar wind streams are  from 12:00\,UT 30 January to 00:00\,UT 4 February in 1995 (stream 1), from 06:00\,UT 17 January to 06:00\,UT 20 January in 2007 (stream 2), from 00:00\,UT 11 February to 12:00\,UT 14 February in 2008 (stream 3), from 12:00\,UT 12 July to 12:00\,UT 15 July in 2008 (stream 4), respectively. The four streams are typical fast streams, with a speed more than 600\,km\,$s^{-1}$, a density roughly 2-4\,cm$^{-3}$, a temperature around 20\,eV, and a magnetic field about 4-6\,nT.

The transformation from $\textrm{CF}_{\textrm{2D}}$ to $\textrm{PSD}_{\textrm{2D}}$ based on the projection-slice theorem \citep{He2013ApJ} is applied to $\textbf{B}$, $\textbf{V}$, $\textbf{Z}_{\textrm{major}}$, and $\textbf{Z}_{\textrm{minor}}$ to calculate $\textrm{PSD}_{\textrm{2D,\textbf{B}}}$, $\textrm{PSD}_{\textrm{2D,\textbf{V}}}$, $\textrm{PSD}_{\textrm{2D},\textbf{Z}_{\textrm{major}}}$, and $\textrm{PSD}_{\textrm{2D},\textbf{Z}_{\textrm{minor}}}$, respectively.
This method yields the relative normalized values ($\textrm{PSD}_{\textrm{2D,relative}}$). To get the absolute values, we use the following formula:
\begin{equation}
\textrm{PSD}_{\textrm{2D,absolute}}\left ( k,\theta_{k}\right)=\textrm{PSD}_{\textrm{2D,relative}}\left ( k,\theta_{k}\right)\cdot \frac{\textrm{Power}_{\textrm{absolute}}}{\textrm{Power}_{\textrm{relative}}}
\end{equation}
with $\textrm{Power}_{\textrm{absolute}}=\int_{f_{0}}^{f_{1}} \textrm{PSD}_{\textrm{FFT}}df$ and $\textrm{Power}_{\textrm{relative}}=\int_{f_{0}}^{f_{1}}\int_{0}^{2\pi} \textrm{PSD}_{\textrm{2D,relative}}\cdot f dfd\theta$. $f_{0}$ and $f_{1}$ stand for the lower and upper limit of the frequency range used to calculate the power, respectively. Here, $f_{0}$ and $f_{1}$ are set to $10^{-4}$ and 0.067~Hz. $\textrm{PSD}_{\textrm{FFT}}$ is obtained by the Fast Fourier transformation of the whole time sequence. Then the residual energy $\textrm{E}_{\textrm{R}}=\textrm{PSD}_{\textrm{2D,absolute,\textbf{V}}}-\textrm{PSD}_{\textrm{2D,absolute,\textbf{b}}}$ , Alfv\'{e}n ratio $\textrm{R}_{\textrm{A}}=\frac{\textrm{PSD}_{\textrm{2D,absolute,\textbf{V}}}}{\textrm{PSD}_{\textrm{2D,absolute,\textbf{b}}}}$ and Els\"asser ratio $\textrm{R}_{\textrm{E}}=\frac{\textrm{PSD}_{\textrm{2D,absolute,\textbf{Z}}_{\textrm{minor}}}}{\textrm{PSD}_{\textrm{2D,absolute,\textbf{Z}}_{\textrm{major}}}}$ in wave-vector space are investigated consequently.

\begin{figure*}
\centering
\includegraphics[width=14cm]{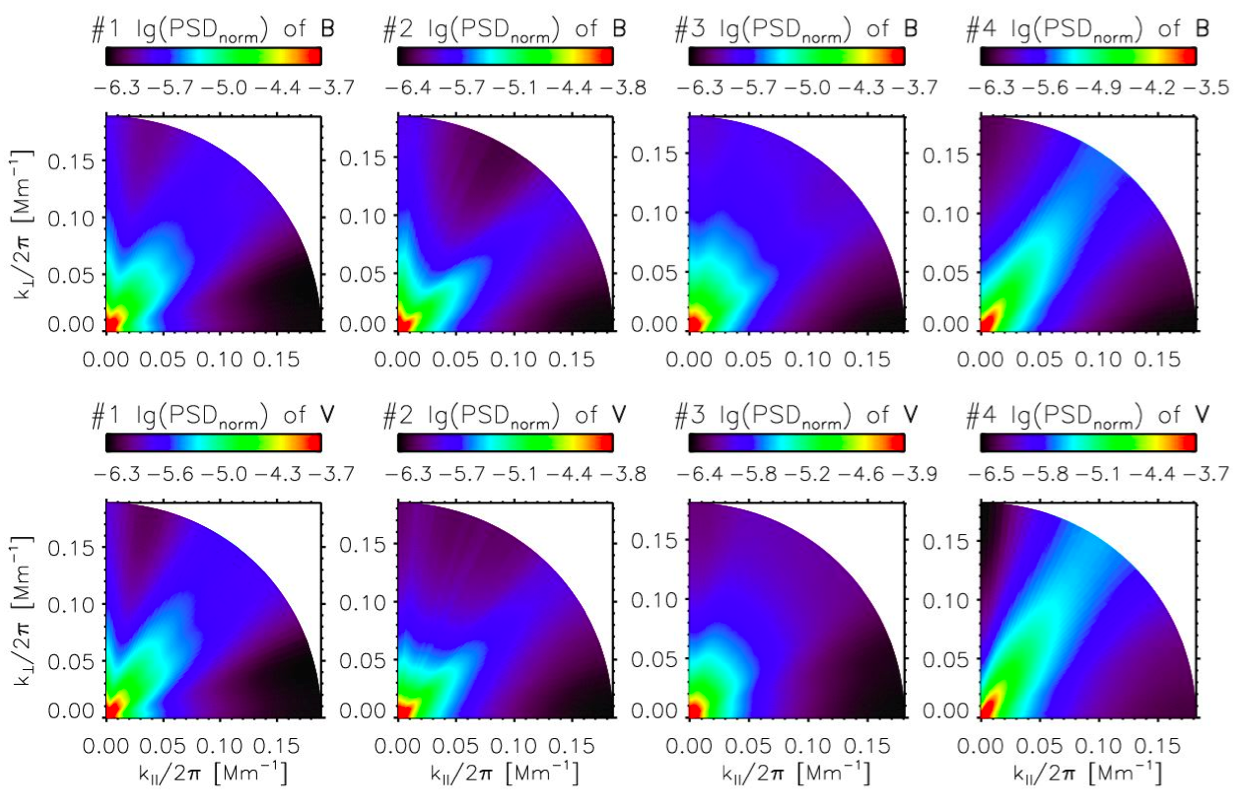}
\caption{The spectra $\textrm{PSD}_{\textrm{2D}}$ of \textbf{B} (upper panels) and \textbf{V} (lower panels), which are normalized to the maximum value, for the four fast solar wind streams.} \label{fig2}
\end{figure*}

Figure~\ref{fig2} displays the spectra $\textrm{PSD}_{\textrm{2D}}$ of magnetic field \textbf{B} (upper panels) and velocity \textbf{V} (lower panels) for the four fast streams. The spectra $\textrm{PSD}_{\textrm{2D}}$ of magnetic field \textbf{B} behave similar to that obtained by \citet{He2013ApJ}. The spectra $\textrm{PSD}_{\textrm{2D}}$ of \textbf{B} for the four fast streams show a similar anisotropic distribution in wave-vector space. The PSD is distributed mainly along a ridge which is inclined toward the $k_{\perp}$ axis. Besides the similarity, the distribution of the spectra $\textrm{PSD}_{\textrm{2D}}$ also shows some difference between different streams. For example, stream 1 and stream 2 show a component which is aligned with the $k_{\perp}$ axis. We are not currently sure whether the difference between different streams is caused by some underlying physical difference or by the method uncertainty or by both. In the future, more effort is needed to be done to quantitatively estimate the method uncertainty and distinguish it from the physical signal. The spectra $\textrm{PSD}_{\textrm{2D}}$ of velocity \textbf{V} show a similar anisotropy pattern as that of magnetic field \textbf{B}, suggesting the signature of oblique Alfv\'{e}n waves.

\begin{figure*}
\centering
\includegraphics[width=14cm]{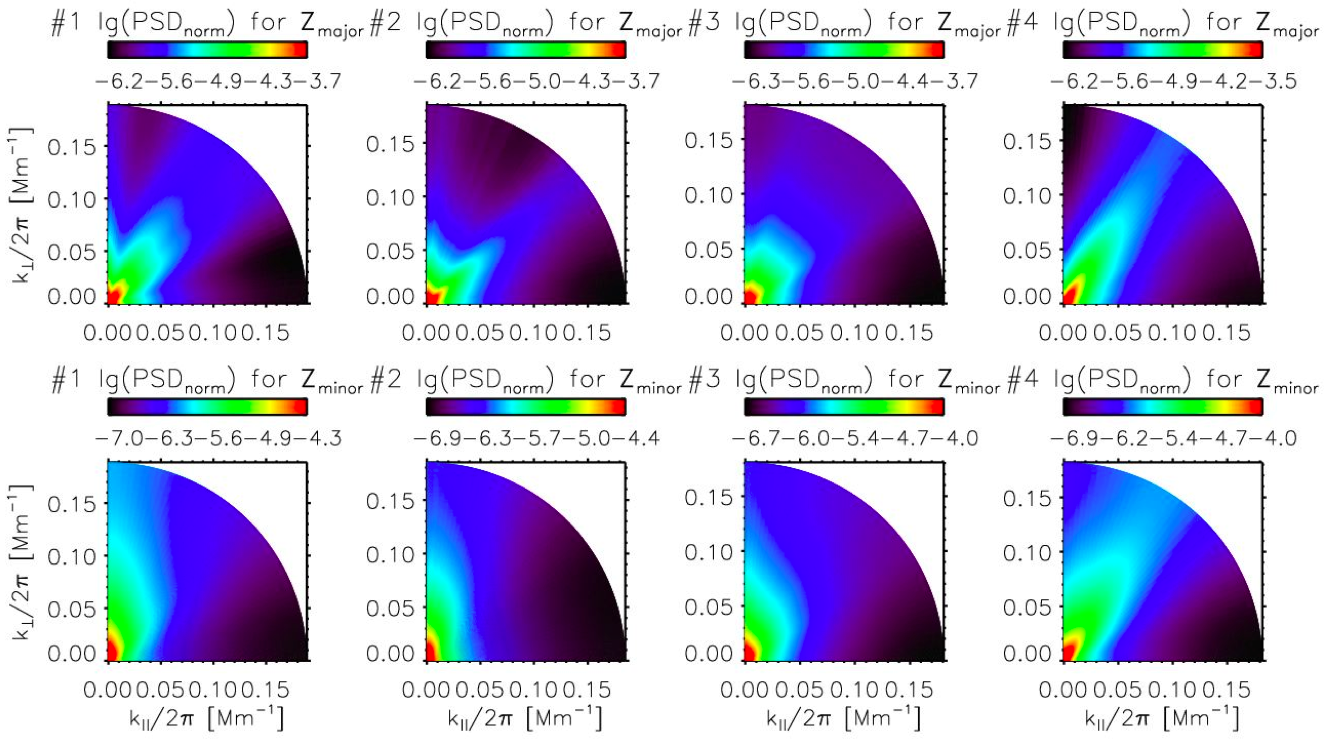}
\caption{The spectra $\textrm{PSD}_{\textrm{2D}}$ of $\textbf{Z}_{\textrm{major}}$ (upper panels) and $\textbf{Z}_{\textrm{minor}}$ (lower panels), which are normalized to the maximum value, for the four fast solar wind streams.} \label{fig3}
\end{figure*}

\begin{figure*}
\centering
\includegraphics[width=14cm]{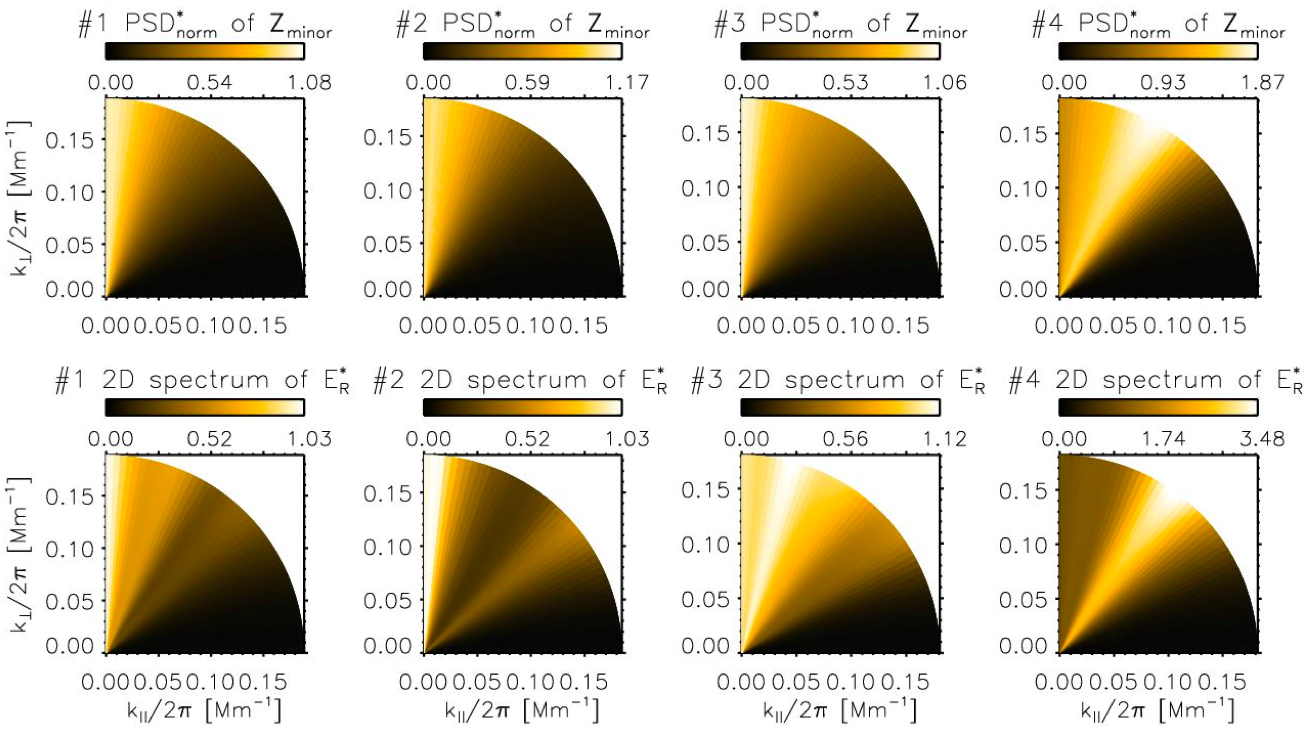}
\caption{The normalized spectra of $\textbf{\textrm{Z}}_{\textrm{minor}}$ ($\frac{\textbf{\textrm{Z}}_{\textrm{minor}}\left(k_{\parallel},k_{\perp}\right)}{\textbf{\textrm{Z}}_{\textrm{minor}}\left(k_{\parallel}=0,k_{\perp}\right)}$;upper panels) and residual energy ($\frac{\textrm{E}_{\textrm{R}}\left(k_{\parallel},k_{\perp}\right)}{\textrm{E}_{R}\left(k_{\parallel}=0,k_{\perp}\right)}$;lower panels), for the four fast solar wind streams.} \label{fig4}
\end{figure*}

To investigate the spectral anisotropy of Els\"asser variables, the Els\"asser spectra in wave-vector space are obtained and shown in Figure~\ref{fig3}. The spectra $\textrm{PSD}_{\textrm{2D}}$ of $\textbf{\textrm{Z}}_{\textrm{major}}$  and $\textbf{\textrm{Z}}_{\textrm{minor}}$ both show anisotropy in the wave-vector space. However, the anisotropy pattern is different for $\textbf{\textrm{Z}}_{\textrm{major}}$ and $\textbf{\textrm{Z}}_{\textrm{minor}}$. The spectra $\textrm{PSD}_{\textrm{2D}}$ of $\textbf{\textrm{Z}}_{\textrm{major}}$ share a similar anisotropic pattern with that of magnetic field \textbf{B}, and velocity \textbf{V}, while  the spectra $\textrm{PSD}_{\textrm{2D}}$ of $\textbf{\textrm{Z}}_{\textrm{minor}}$ show a very different anisotropy with the main features of $\textrm{PSD}_{\textrm{2D}}$ distributed along the $k_{\perp}$ axis. The spectra $\textrm{PSD}_{\textrm{2D}}$ of $\textbf{\textrm{Z}}_{\textrm{minor}}$ normalized to the $\textrm{PSD}_{\textrm{2D}}$ with the same $k_{\perp}$ but with $k_{\parallel}=0$ (upper panels in Figure~\ref{fig4}) reveal further evidence that the spectra $\textrm{PSD}_{\textrm{2D}}$ of $\textbf{\textrm{Z}}_{\textrm{minor}}$ is mainly distributed at small $k_{\parallel}$. These results suggest that the anisotropy of the subdominant mode $\textbf{\textrm{Z}}_{\textrm{minor}}$ is stronger than that of the dominant mode $\textbf{\textrm{Z}}_{\textrm{major}}$, which is consistent with the observational result based on the reduced spectrum \citep{Wicks2011PhRvL} and the simulation result \citep{Cho2014ApJ}.

The residual energy $\textrm{E}_{\textrm{R}}$ for all the four fast streams is less than 0, meaning that the magnetic energy is dominant over kinetic energy. The residual energy $\textrm{E}_{\textrm{R}}$ is normalized to $k_{\perp}$ axis, using the formula $\textrm{E}_{\textrm{R,norm}}=\frac{\textrm{E}_{\textrm{R}}\left(k_{\parallel},k_{\perp}\right)}{\textrm{E}_{R}\left(k_{\parallel}=0,k_{\perp}\right)}$, and shown in the lower panels in Figure~\ref{fig4}. As seen from the normalized residual energy $\textrm{E}_{\textrm{R,norm}}$, the residual energy $\textrm{E}_{\textrm{R}}$ is concentrated at small $k_{\parallel}$. This result gives the clear observational support to the simulation results of \citet{Boldyrev2009PhRvL} and \citet{Wang2011ApJ}, which showed a condensate of magnetic energy during cascading of Alf\'ven waves due to the breakdown of the mirror symmetry in nonbalanced turbulence.

\begin{figure*}
\centering
\includegraphics[width=14cm]{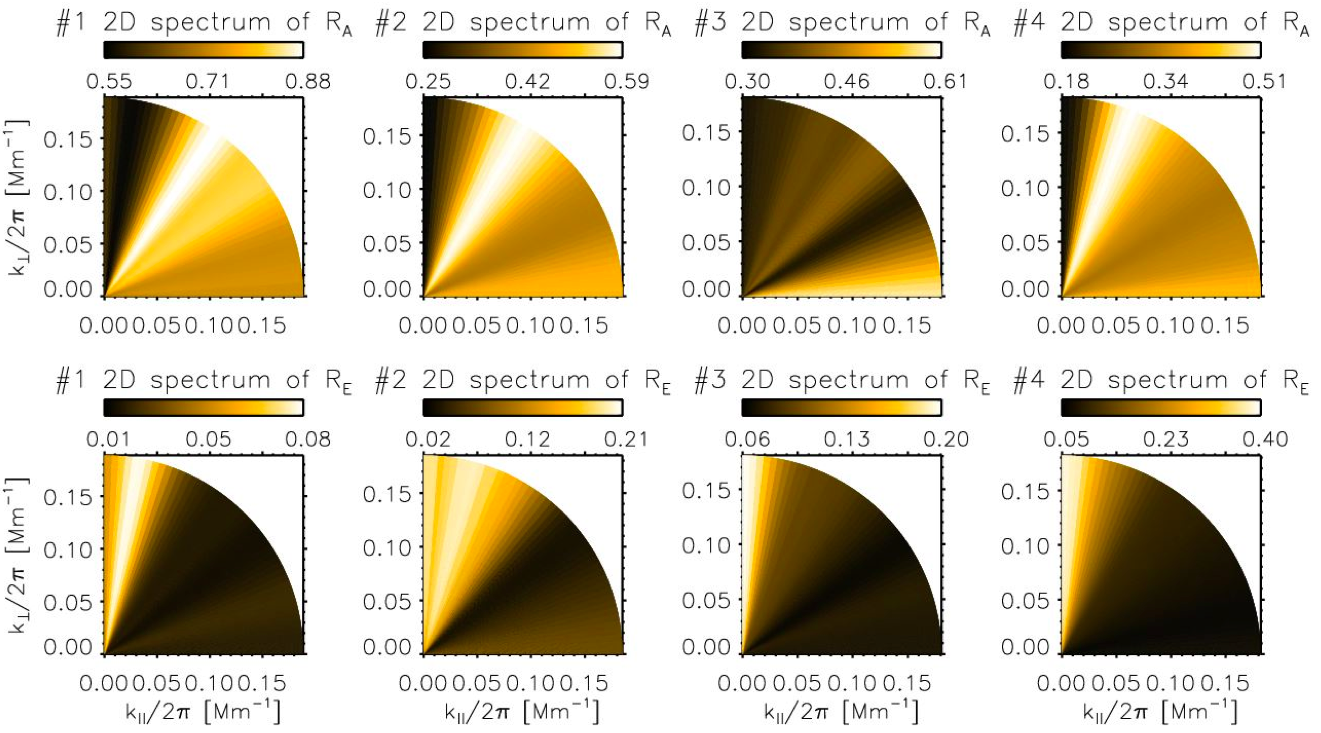}
\caption{The wave-vector distribution of Alfv\'en ratio $\textrm{R}_{\textrm{A}}$ (upper panels) and Els\"asser ratio $\textrm{R}_{\textrm{E}}$ (lower panels) for the four fast solar wind streams.} \label{fig5}
\end{figure*}

The distribution of $\textrm{R}_{\textrm{A}}$ (upper panels in Figure~\ref{fig5}) and $\textrm{R}_{\textrm{E}}$ (lower panels in Figure~\ref{fig5}) both show anisotropy. Close to the $k_{\perp}$ axis, $\textrm{R}_{\textrm{A}}$ becomes smaller, suggesting that the dominance of magnetic energy over the kinetic energy becomes significant at small $k_{\parallel}$. $\textrm{R}_{\textrm{E}}$ close to the $k_{\perp}$ axis is much larger than at other angles, suggesting that the difference between  the energy of $\textbf{\textrm{Z}}_{\textrm{major}}$ and that of $\textbf{\textrm{Z}}_{\textrm{minor}}$ is larger close to the $k_{\perp}$ axis.

\section{Summary and discussions}
In this paper, we first did a benchmark test of the conversion between $\textrm{CF}_{\textrm{2D}}$ and $\textrm{PSD}_{\textrm{2D}}$, confirming that the conversion obtained directly from the corresponding formula, by the transformation based on the projection-slice theorem, and by the transformation based on two dimensional inverse Fourier are in accordance with each other. This experiment manifests the applicability of the transformation based on the projection-slice theorem to estimate $\textrm{PSD}_{\textrm{2D}}$.

Based on the transformation, we investigated the spectral anisotropy of Els\"asser variables in 2D wave-vector space for the first time. We also studied the distribution of residual energy $\textrm{E}_{\textrm{R}}$, Alfv\'en ratio $\textrm{R}_{\textrm{A}}$ and Els\"asser ratio $\textrm{R}_{\textrm{E}}$, which have not been studied in the $(k_{\parallel},k_{\perp} )$ space before. Four fast streams observed by the WIND spacecraft were studied in this work.

The spectra $\textrm{PSD}_{\textrm{2D}}$ of $\textbf{Z}_{\textrm{major}}$ and $\textbf{Z}_{\textrm{minor}}$ both show anisotropy in the wave-vector space. However, the anisotropic pattern of $\textbf{Z}_{\textrm{major}}$ and $\textbf{Z}_{\textrm{minor}}$ is different and the anisotropy of $\textbf{Z}_{\textrm{minor}}$ seems stronger than that of $\textbf{Z}_{\textrm{major}}$, which is consistent with the observational result from the reduced spectrum \citep{Wicks2011PhRvL} and the simulation result \citep{Cho2014ApJ}.

 For each of the four fast streams, the spectra $\textrm{PSD}_{\textrm{2D}}$ of \textbf{B}, \textbf{V}, and $\textbf{Z}_{\textrm{major}}$ share a similar anisotropic pattern as that obtained by \citet{He2013ApJ}, The spectra $\textrm{PSD}_{\textrm{2D}}$ is distributed mainly along a ridge which is inclined toward the $k_{\perp}$ axis. This suggests that $\textbf{Z}_{\textrm{major}}$ probably correspond to the oblique Alfv\'{e}nic fluctuations propagating outwardly.

Differently from that of \textbf{B}, \textbf{V}, and $\textbf{Z}_{\textrm{major}}$, the spectra $\textrm{PSD}_{\textrm{2D}}$ of $\textbf{\textrm{Z}}_{\textrm{minor}}$ is distributed mainly along the $k_{\perp}$ axis. The Els\"asser ratio $\textrm{R}_{\textrm{E}}$  is larger at large $\theta_{kB}$ angles than at other angles, suggesting that the difference between the spectra $\textrm{PSD}_{\textrm{2D}}$ of $\textbf{\textrm{Z}}_{\textrm{major}}$ and that of $\textbf{\textrm{Z}}_{\textrm{minor}}$ becomes more evident when it is getting close to the $k_{\perp}$ axis. The spectra $\textrm{PSD}_{\textrm{2D}}$ of $\textbf{\textrm{Z}}_{\textrm{minor}}$ normalized to the $\textrm{PSD}_{\textrm{2D}}$ with the same $k_{\perp}$ but with $k_{\parallel}=0$ further demonstrates that the power of $\textbf{\textrm{Z}}_{\textrm{minor}}$ is concentrated at small $k_{\parallel}$. The Alfv\'en ratio $\textrm{R}_{\textrm{A}}$ close to the $k_{\perp}$ axis is much smaller compared to that at other angles. If the plasma is thermally anisotropic and component-drifted, the Alfv\'en ratio will be very low, even when $\textbf{\textrm{Z}}_{\textrm{minor}}$ stands for the inward propagating Alfv\'en wave. So, the presence of inward propagating Alfv\'en waves can not be excluded. If the cascade of $\textbf{\textrm{Z}}_{\textrm{minor}}$ is driven by $\textbf{\textrm{Z}}_{\textrm{major}}$, this may suggest that the cascade is anisotropic and probably mainly along the $k_{\perp}$ direction. The magnetic structure without velocity fluctuations and the non-Alfv\'enic fluctuation with $k_{\parallel}=0$ both could lead to the power concentration of $\textbf{\textrm{Z}}_{\textrm{minor}}$ and the low Alfv\'en ratio. Further work is required in the future to understand what $\textbf{\textrm{Z}}_{\textrm{minor}}$ mostly represents.

 Though the spectra of \textbf{B} and \textbf{V} share a similar spectral anisotropic pattern, there are still differences between them as revealed by the anisotropic distribution of $\textrm{E}_{\textrm{R}}$ and $\textrm{R}_{\textrm{A}}$. Close to the $k_{\perp}$ axis, $\textrm{R}_{\textrm{A}}$ becomes smaller and $\left|\textrm{E}_{\textrm{R}}\right|$ becomes larger, suggesting that the dominance of the magnetic energy over the kinetic energy becomes significant. The residual energy condensate at small $k_{\parallel}$ confirms observationally the findings in the simulation results of \citet{Boldyrev2009PhRvL} and \citet{Wang2011ApJ}.

It should be noted that, $\textbf{\textrm{Z}}_{\textrm{minor}}$ may be anti-correlated with $\textbf{\textrm{Z}}_{\textrm{major}}$ due to the dominance of magnetic energy over kinetic energy. The unequipartition between magnetic and kinetic energy may be the case for Alfv\'en waves with kinetic effects if the plasma is thermally anisotropic and component-drifted. We have tried to re-estimate the spectra $\textrm{PSD}_{\textrm{2D}}$ of $\textbf{\textrm{Z}}_{\textrm{major}}$ and $\textbf{\textrm{Z}}_{\textrm{minor}}$ after correcting for kinetic effects from the thermal anisotropy. The recalculated distribution of $\textrm{PSD}_{\textrm{2D}}$ of $\textbf{\textrm{Z}}_{\textrm{major}}$ remain almost unchanged. However, the $\textrm{PSD}_{\textrm{2D}}$ of $\textbf{\textrm{Z}}_{\textrm{minor}}$ after correcting the thermal anisotropy can not be reconstructed with good quality, which might be due to the possible over-correction of the thermal anisotropy on the weak signal of $\textbf{\textrm{Z}}_{\textrm{minor}}$.

In the critical balance theory of \citet{GS1995ApJ}, the eddies are filament shaped. In simulations the eddies usually have a ribbon shape \citep{Muller2000PhRvL,Biskamp2000PhPl,Maron2001ApJ} rather than a filament. \citet{Boldyrev2006PhRvL} extended critical balance theory to account for this 3D anisotropy. \citet{Chen2013ApJ} investigated the local three dimensional structure functions of the inertial range plasma turbulence based on observation for the first time. They found that the Alfv\'enic fluctuations are three-dimensionally anisotropic dependent on the scales. In future, we intend to extend this method to three dimension to investigate the PSD in 3D wave-vector space and compare the result with former theoretical and simulation results. To promote the usage of the method in the community, further calibration of this method , e.g. to compare the reconstructed PSD with the known PSD, the turbulent fluctuations of which is measured for the reconstruction \citep{Horaites2015PhRvL}, is required \citep{Oughton2015rsta} (private communication with Tulasi Parashar).

\acknowledgments{The group from Peking University was supported by NSFC under 41174148, 41222032, 41231069, 41421003, 41474147, 41274172, and 41474148. J.S.H., C.Y.T., C.H.K.C., X.W., and R.W. are also members of the ISSI/ISSI-BJ international team 304.}


\end{document}